\documentclass[a4paper, 12pt]{article}
\usepackage[T1]{fontenc}
\usepackage[utf8]{inputenc}
\usepackage{amstext,amsmath,amssymb}
\usepackage{graphicx,epstopdf}
\usepackage{authblk}
\usepackage{hyperref}
\usepackage[skip=5pt]{caption}

\title{A new thought experiment on relativistic length contraction}
\author[1]{Biplab Raychaudhuri \thanks{biplabphy@visva-bharati.ac.in}}
\author[2]{Souvik Ghose \thanks{dr.souvikghose@gmail.com}}
\author[2]{Arunabha Bhadra \thanks{aru\_bhadra@yahoo.com}}
\affil[1]{Department of Physics, Visva-Bharati University, Santiniketan, West Bengal, India, 731235}
\affil[2]{High Energy \& Cosmic Ray Research Center, North Bengal University, Siliguri, Darjeeling, West Bengal, India, 734013}
\begin{document}
\maketitle
\begin{abstract}
Relativistic length contraction is revisited and a simple but new thought experiment is proposed in which an apparent asymmetric situation is developed between two different inertial frames regarding detection of light that comes from a chamber to an adjacent chamber through a movable slit. The proposed experiment does not involve gravity, rigidity or any other dynamical aspect apart from the kinematics of relative motion; neither does it involve any kind of non-uniformity in motion. The resolution of the seemingly paradoxical situation has finally been discussed. 
\end{abstract}

\section{Introduction} 
Special theory of relativity (STR), which assumes Lorentz invariance (LI) as a fundamental symmetry of nature, has a central place in modern physics. According to one of the two postulates of STR, the speed of light in vacuum, unlike that of any other particle, remains invariant to any inertial observer independent of the motion of the source or the observer~\cite{einstein2003meaning}. This idea of constant light speed is so  counter intuitive that since the formulation of the theory several researchers have critically examined its consistency and explored different surprising consequences of the Lorentz invariance, particularly in search of a real paradox in the theory. Length contraction and time dilation as well as their apparently paradoxical consequences are discussed in introductory textbooks to familiarize  new students with the curious consequences of the postulates of STR \cite{halliday2013fundamentals}. Several such incongruities were proposed and subsequently resolved in the literature. Resolution of most of these alleged paradoxes, if not all, involves the presence of an external force, rigidity, or some other subtle issues. While the most agreed upon view on length-contraction and time dilation is that the effects are real and can be demonstrated through experiment \cite{bailey1977measurements}, some people have remained skeptical \cite{jefimenko1998experimental}. Nevertheless, the thought experiments on the consequences of STR have retained their importance, more so from the pedagogical view point.
 In the present work we propose a new variant of the length contraction paradox (and also advance a resolution of it) which can be discussed based on the kinematic aspects of special relativistic effects only. While, like all the other length contraction gedanken in the literature (see Appendix), the present one does not advance any new idea, it would, hopefully, help students understand the curious aspect of special theory of relativity without exposing them to the more confounding issues which can be introduced at a much later stage.

\section{An apparent paradox}
\begin{figure}
\centering
\includegraphics[width=\linewidth]{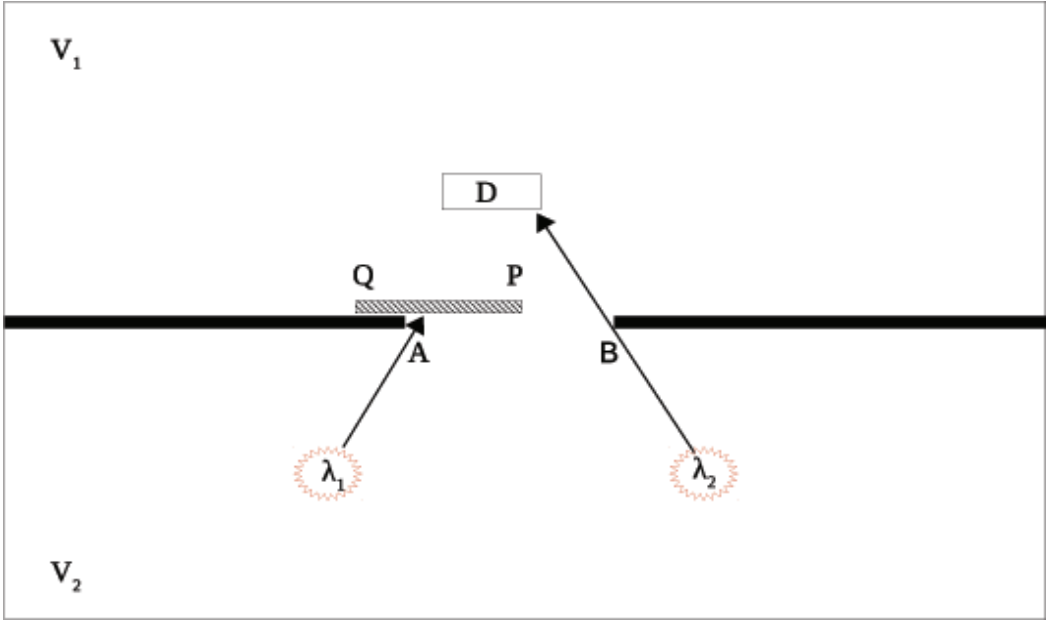}
\caption{From laboratory frame: detector always receives photon.}
\label{lcp1}
\end{figure}

\begin{figure}
\centering
\includegraphics[width=\linewidth]{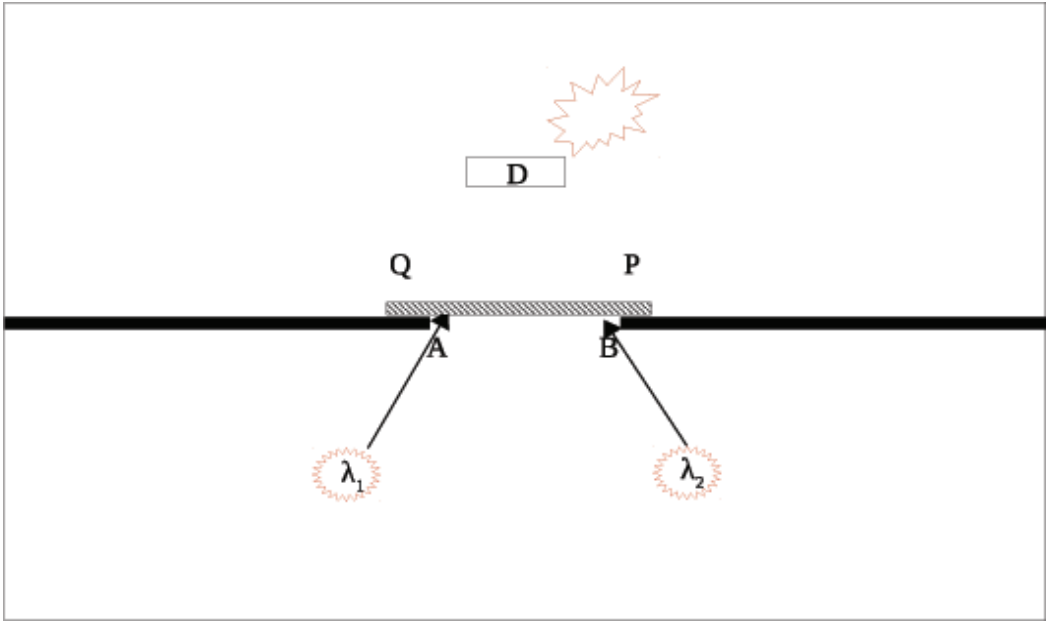}
\caption{From rod-bound frame: both sources are blocked for some time.}
\label{lcp2}
\end{figure}

Let us consider two  adjacent chambers $V_1$ and $V_2$ as shown in Fig~\ref{lcp1}. A slit $AB$ of length $L$ connects $V_1$ and $V_2$. There are two identical light sources $\lambda_1$ and $\lambda_2$ in chamber $V_2$, just infinitesimal distance below the points $A$ and $B$.  A $100\%$ efficient photon detector $D$ sits in $V_1$ and  infinitesimally above the middle point of the slit.  As long as the slit is open the photons from $\lambda_1$ and $\lambda_2$  pass grazing the edge of the slit and are detected at $D$. The detector, however, is equipped with an explosive device which would be detonated if at any moment the detector count fell to zero. There is also a bar $PQ$ of rest length $L$ that is moving freely with relativistic speed along the boundary of two chambers towards positive $X$-axis as per an observer $O$ who is at rest with respect to the two chambers, slit, light sources and detector. Let us associate a frame of reference $S(t, x, y, z$) with the observer $O$ and a frame of reference $S'(t', x', y', z')$ ($c=1$) with an observer $O'$ which is at rest with the bar $PQ$. The axes of both  frames coincide at the point $A$ at a certain time to which we  assign the origin of the two frames $S$ and $S'$. Thus the  $X$ coordinates of the points $A$ and $B$ are $0$ and $x_B$ respectively in frame $S$ and $0$ and $x_B'$ in frame $S'$.
Let us first consider the observation of the observer $O$.  The bar is moving towards the slit from its edge $A$ in its frame.  Initially the front end of the bar $P$ coincides with $A$. As the bar moves towards $B$, the light from the source $\lambda_1$ will be blocked; but the detector $D$ will continue receiving light from $\lambda_2$. At some instant $t_1$ the back end ($Q$) of the bar will reach the point $A$. The front end $P$ of the bar is then, say, at point $x_1 $. The length of the bar in frame $S$ is $x_1$. Due to length contraction, the length of the bar in the frame $S$ is $L/\gamma$, where $\gamma=1/\sqrt{(1-v^2/c^2 )}$. So $x_1= L/\gamma < L$, which implies that a part of the slit is still open and the detector $D$ is receiving light from $\lambda_2$. Next, consider the situation when the front end $P$ of the bar reaches $B$ (at $x_B$) at a time $t_2$. At that instant the back end $Q$ of the bar is at  $x_2$ and the length of the bar in frame $S$ is $x_B - x_2$. The length contraction implies $x_2> 0$. So light from the source $\lambda_2$ is now blocked to reach the detector $D$, but light from the source $\lambda_1$  still reaches $D$. Thus, the observer in frame $S$ concludes that the detector $D$ is continuously receiving light.
Now, let us consider the experiment in the frame of the observer $O'$ (Fig.~\ref{lcp2}). She observes that the slit, light sources and the detectors are moving towards the bar in the $-$ve $X$ direction after $t'=0$.

At time ${t'}_1$ the far end $B$ of the slit will reach point $P$ and at the same instant the point $A$ will reach, say, $x'_1$. The length of the slit, according to the observer $S'$, will be $x'_1$ . Due to length contraction $x'_1= L/\gamma < L$. Since the slit is now shorter than the bar, no light reaches the detector from any of the two sources, $\lambda_1$ or $\lambda_2$.  This situation will continue till the point $A$ of the slit reaches the point $Q$ of the bar.  The observer $O'$ then definitely sees an explosion! 

\section{Resolution of the proposed paradox}

So far we have concentrated on passing of light through the slit and not on detection of light by the detector. The detector is at rest with respect to the observer $O$ but is in motion according to the observer $O^{\prime}$. The light will take exactly the same time while propagating from the point $A$ to $D$ and the point $B$ to $D$ in frame $O$. But the same is not true in frame $O^{\prime}$. In the view point of $O^{\prime}$ after the point B coincides with the point P light from the sources $\lambda_1$ and $\lambda_2$ cannot reach the upper chamber till the point $A$ reaches $Q$. So we need to calculate the time difference in frame $O^{\prime}$ between the propagation of light from the point $B$, when it coincides with $P$, to $D$ and the point $A$, as it arrives at $Q$, to $D$. 

Let $t_{BD}^{\prime}$ is the time taken by a photon to reach the detector $D$ from the point $B$ as $B$ coincides with $P$. The vertical distance between B/A and D is say $y$ and the horizontal distance (in $x^{\prime}$ coordinates) between B/A and D is $ \frac{L}{2\gamma }$. Therefore, one can write
\begin{equation}
\sqrt{\left(\frac{L}{2\gamma} + v t_{BD}^{\prime} \right)^{2} + y^2} = ct_{BD}^{\prime}
\end{equation}
which leads to
\begin{equation}
t_{BD}^{\prime} = \frac{\frac{Lv}{\gamma} \pm \sqrt{\frac{L^2 v^2}{\gamma^2} + 4 \left(c^2-v^2\right) \left(\frac{L^2}{4\gamma^2} +y^2 \right)}}{2\left(c^2-v^2\right)}
\end{equation}
Since $t_{BD}^{\prime}$ has to be positive, we have to take only positive root of the above solution. 
Similarly,
\begin{equation}
t_{AD}^{\prime} = \frac{-\frac{Lv}{\gamma} \pm \sqrt{\frac{L^2 v^2}{\gamma^2} + 4 \left(c^2-v^2\right) \left[\frac{L^2}{4\gamma^2} +y^2 \right]}}{2\left(c^2-v^2\right)}    
\end{equation}
Thus the difference in propagation time of photon from $B$ to $D$ and $A$ to $D$ $\Delta t_{prop}^{\prime} = \frac{L\gamma v}{c^2}$ which is interestingly independent of y i.e. vertical position of the detector. 


In the primed frame (rest frame of the rod) the last photon that escapes the rod, when $B$ coincides with $P$, reaches $D$ at $t^{\prime}_{D1}=t^{\prime}_1 + t_{BD}^{\prime}$. When $A$ reaches $Q$, the first photon that escapes, reaches $D$ at $t^{\prime}_{D2} =t^{\prime}_2 +t_{AD}^{\prime}$. The time difference in arrival of these two photons at $D$ is $\Delta t^{\prime}_{D}= t^{\prime}_{D1}-t^{\prime}_{D2}$  
\begin{equation}
\Delta t^{\prime}_{D} =\frac{L}{v} \left(\gamma -1 \right)    
\end{equation}
which is positive. Thus the the first photon that enters the upper chamber when $A$ coincides $Q$ reaches the detector $D$ before the last photon that enters the upper chamber when $B$ coincides $P$. This implies that even in the primed frame the detector $D$ continuously receives photons and manages to avoid an explosion.


\section{Conclusion}
The above puzzle relies only on the length contraction phenomenon. Its formation and resolution do not involve any external force; neither does it involve any non-uniform motion. The seemingly paradoxical situations in STR arise from the counter intuitive nature of the time dilation and the length contraction. In the present work an apparent  inconsistency  is resolved by taking the leverage of using length contraction in one frame and time dilation in the other (i.e. essentially using the invariance of four dimensional length element ($ds^2$)) like in the famous muon decay problem. The (apparent) paradoxes involving pure time dilation or pure length contraction are rather difficult to resolve in a straight forward way. The resolution of twin paradox problem orients around acceleration of one of the twin while the resolution of the pole and garage paradox relies on rigidity/stress of pole material. STR is a very successful physical theory. Not only the theory has been found to explain a wide range of experimental results but it has also been well tested experimentally with an enormous range and diversity, and the agreement between theory and experiment is excellent. Many of the predictions of the theory are at times counter-intuitive and over the years researchers have explored many enigmatic situations. However, subsequent resolutions of these (apparent) paradoxes only re-establish the great inner consistency of the theory. In this ambit the present work seeks to serve some pedagogical purpose by inviting a student to the curious world of the STR. 

\section*{Acknowledgement}
Authors are grateful to Professor R. H. Price for pointing the direction of possible resolution of the proposed paradox. BR likes to thank Inter-University Centre for Astronomy and Astrophysics (IUCAA), Pune for their hospitality and facilities they extended towards him during his visits under their Visiting Associateship programme.

\section* {Appendix}
Historically the concept of length contraction of moving objects in the direction of motion was introduced independently in an ad hoc manner by FitzGerald~\cite{fitzgerald2001ether} and Lorentz~\cite{lorentz1892relative} in order to explain the negative results of the Michelson-Morley experiment. It was realized later that length contraction is an inevitable consequence of Lorentz transformation. This effect also leads to many puzzling situation many of which found their way into standard special relativity text books. Some of the well-known (apparent) paradoxes are mentioned below.\\
\textbf{Rod and hole problem:}\\ 
A rod, moving on a smooth table, approaches a hole of the same dimension \cite{rindler1961length,rindler1961erratum,shaw1962length}. For the observer on the rod the hole is Lorentz contacted and the rod avoids falling through it. For an observer inside the hole the rod is length-contracted and should fall through it. The paradox is resolved challenging the validity of the rigid-rod assumption within relativistic framework (for more on rigidity see \cite{pierce2007lock}) . The force of gravity is also present to fecilitate the fall in the first place. This paradox was latter modifid to avoid the force of gravity \cite{marx1967lorentz}. In the new scenario both the rod and the hole moves in mutually perpendicular direction and happens to be at the same place at some instant. The paradox is resolved showing that the rod would appear to be titled in one frame.\\ 
\textbf{The pole and barn/garage paradox:}\\ 
Here \cite{dewan1963stress}, a person runs into a garage with a horizontal pole, the pole and the garage being of the same length. From the perspective of the runner the garage is length contracted and the pole can not be fitted within. For a garage observer, on the other hand, the pole is shortened and should easily fit inside it. Consequently, in one frame both the garage doors can be closed with the pole trapped inside but in the other frame that is simply not possible. Once again the paradox is resolved based on an argument that the relativistic pole is no longer rigid in the non-relativistic sense and the back end of it can not know instantly when the front end hits the garage door.\\  
\textbf{Paradox with an electrical circuit:}\\
Sastri \cite{sastry1987length} proposed an interesting paradox with a simple electrical circuit which is completed by a moving shuttle. Consequently a bulb is turn on. From the frame of the circuit the length of the shuttle is shorter than the gap in the circuit and it fails to connect the two ends. From the frame of the shuttle however, the gap is shorter and successfully bridged by the shuttle for an instant, turning on the light. Although successfully resolved this paradox brings in the propagation of electrical signal into consideration.


\begin{thebibliography}{10}

\bibitem{einstein2003meaning}
A.~Einstein, {\em The meaning of relativity} (Routledge, 2003).

\bibitem{halliday2013fundamentals}
D.~Halliday, R.~Resnick and J.~Walker, {\em Fundamentals of physics} (John
  Wiley \& Sons, 2013).

\bibitem{bailey1977measurements}
J.~Bailey, K.~Borer, F.~Combley, H.~Drumm, F.~Krienen, F.~Lange, E.~Picasso,
  W.~Von~Ruden, F.~Farley, J.~Field {\em et~al.}, {\em Nature} {\bf 268}, 301
  (1977).

\bibitem{jefimenko1998experimental}
O.~D. Jefimenko, {\em Zeitschrift f{\"u}r Naturforschung A} {\bf 53}, 977
  (1998).

\bibitem{fitzgerald2001ether}
G.~FitzGerald, {\em Lorentz And Poincare Invariance: 100 Years Of Relativity}
  {\bf 8}, p.~25  (2001).

\bibitem{lorentz1892relative}
H.~A. Lorentz, {\em Zittingsverlag Akad. V. Wet} {\bf 1}, 74  (1892).

\bibitem{rindler1961length}
W.~Rindler, {\em American Journal of Physics} {\bf 29}, 365  (1961).

\bibitem{rindler1961erratum}
W.~Rindler, {\em American Journal of Physics} {\bf 29}, 859  (1961).

\bibitem{shaw1962length}
R.~Shaw, {\em American Journal of Physics} {\bf 30}, 72  (1962).

\bibitem{pierce2007lock}
E.~Pierce, {\em American Journal of Physics} {\bf 75}, 610  (2007).

\bibitem{marx1967lorentz}
E.~Marx, {\em American Journal of Physics} {\bf 35}, 1127  (1967).

\bibitem{dewan1963stress}
E.~M. Dewan, {\em American Journal of Physics} {\bf 31}, 383  (1963).

\bibitem{sastry1987length}
G.~Sastry, {\em American Journal of Physics} {\bf 55}, 943  (1987).

\end{thebibliography}

\end{document}